\documentclass{article}

\PassOptionsToPackage{numbers, compress}{natbib}
\usepackage[preprint]{neurips_2025}
\usepackage{amsmath}

\usepackage[utf8]{inputenc} 
\usepackage[T1]{fontenc}    
\usepackage{hyperref}       
\usepackage{url}            
\usepackage{booktabs}       
\usepackage{amsfonts}       
\usepackage{nicefrac}       
\usepackage{microtype}      
\usepackage{xcolor}         
\usepackage{graphicx} 
\usepackage{hyperref}       
\usepackage{url}            
\usepackage{booktabs}       
\usepackage{amsfonts}       
\usepackage{nicefrac}       
\usepackage{microtype}      
\usepackage{xcolor}         

\usepackage{caption} 
\usepackage{graphicx}
\usepackage{booktabs} 
\usepackage{amssymb}
\usepackage{amsmath}
\usepackage{multirow}
\usepackage{subcaption}
\usepackage{float}
\usepackage{mathtools}
\usepackage{bbold}

\usepackage[linesnumbered,ruled,vlined]{algorithm2e}
\DontPrintSemicolon

\SetKwComment{Comment}{\color{green!50!black}// }{}

\SetKwProg{Function}{function}{}{}

\usepackage{pythontex} 
\usepackage{makecell}
\usepackage{wrapfig}

\title{Hybrid Mamba–Transformer Decoder for Error-Correcting Codes}

\author{%
  Shy-el Cohen \\
  School of Electrical and Computer Engineering \\
  Ben-Gurion University of the Negev \\
  \texttt{shyelc@post.bgu.ac.il} \\
  \And
  Yoni Choukroun \\
  Blavatnik School of Computer Science \\
  Tel Aviv University \\
  \texttt{choukroun.yoni@gmail.com} \\
  \And
  Eliya Nachmani \\
  School of Electrical and Computer Engineering \\
  Ben-Gurion University of the Negev \\
  \texttt{eliyanac@bgu.ac.il} \\
}

\begin{document}

\maketitle

\maketitle

\begin{abstract}
We introduce a novel deep learning method for decoding error correction codes based on the Mamba architecture, enhanced with Transformer layers. Our approach proposes a hybrid decoder that leverages Mamba’s efficient sequential modeling while maintaining the global context capabilities of Transformers. To further improve performance, we design a novel layer-wise masking strategy applied to each Mamba layer, allowing selective attention to relevant code features at different depths. Additionally, we introduce a progressive layer-wise loss, supervising the network at intermediate stages and promoting robust feature extraction throughout the decoding process. Comprehensive experiments across a range of linear codes demonstrate that our method significantly outperforms Transformer-only decoders and standard Mamba models.
\end{abstract}
\section{Introduction}

Deep learning–based decoders have achieved remarkable success in decoding error-correcting codes in recent years. Notable examples include Neural Belief Propagation \citep{nachmani2018deep}, which learns weights of the message-passing algorithm; Neural Min-Sum \citep{Lugosch_2017,9427170}, which approximates the classical min-sum decoder with trainable parameters; Neural RNN decoder \citep{kim2018communication} for convolutional and turbo error correcting codes. Recently, diffusion-based decoders \citep{choukroun2022denoisingdiffusionerrorcorrection}, which model channel noise as a diffusion process that can be reversed; and Transformer-based decoders \citep{choukroun2022errorcorrectioncodetransformer,choukroun2024foundation, park2024crossmpt,zhengwhite}, which exploit self-attention to capture the code structure, reached state-of-the-art performance in neural decoding. However, despite their individual strengths, these methods either incur a high computational cost, compared to classical decoders, or fail to achieve state-of-the-art performance on some codes.

In this work, we propose a novel hybrid decoder that combines the Mamba architecture \citep{gu2023mamba} - known for its highly efficient sequential modeling and low runtime latency - with transformer layers \citep{vaswani2017attention} that provide global receptive fields throughout the codeword. Concretely, we introduce a layer-wise masking strategy within each Mamba block, enabling the model to selectively focus on the most informative subsets of bits at varying depths. To further bolster the training dynamics, we propose a layer-wise loss that provides intermediate supervision at each decoding stage. This auxiliary loss not only promotes better gradient propagation through deep networks but also encourages the extraction of the decoding codeword at each stage enabling intermediate estimation of the decode codeword. 

Extensive experiments on a diverse suite of binary linear block codes, including BCH, Polar, and LDPC codes, demonstrate that our Mamba–Transformer decoder consistently surpasses both Transformer-only decoders and conventional Mamba implementations. We report relative improvements of up to $18\%$ in BER for BCH and Polar codes, and on-par on LDPC codes, while improving inference speed compared to previous works.

\section{Related works}
\subsection{Neural Decoders}
In recent years, the study of deep learning-based decoders for error correction codes has emerged as a vibrant and rapidly evolving research area \citep{gruber2017deep}.  Two broad paradigms have been pursued: model-based architectures, which embed the structure of classical decoding algorithms into neural networks, and model-free architectures, which treat decoding as a purely data-driven mapping.
\paragraph{Model‐based neural decoders}  
In model-based approaches, the computational graph of a traditional message-passing decoder is reinterpreted as a deep network with trainable parameters.  Neural Belief Propagation (NBP) first demonstrated this idea by assigning learnable weights to the edges and messages of the belief propagation algorithm, resulting in a decoder that jointly optimizes its update rules through gradient‐based training \citep{nachmani2016learning}.  Building on NBP, the Neural Min-Sum decoder approximates the classical Min-Sum algorithm by introducing scalar and vector weight parameters into its summation and normalization steps \citep{Lugosch_2017,dai2021learning,kwak2022neural,kwak2023boosting}.  This parameterization retains the low-complexity structure of Min-Sum while achieving performance on par with more expensive decoders.  To further reduce inference cost, pruning techniques have been applied to compress these networks, systematically removing redundant connections and yielding lightweight variants without significant performance degradation \citep{buchberger2020pruning}.
\paragraph{Model-free neural decoders}  
In contrast, model-free decoders rely solely on the representational power of generic neural architectures.  Early work employed fully-connected networks to directly map noisy codewords to their nearest valid codewords, demonstrating feasibility on short block codes \citep{cammerer2017scaling}.  Subsequent studies showed that such networks can scale to moderate block lengths without overfitting \citep{bennatan2018deep}.  More advanced generative frameworks have also been introduced: diffusion‐based decoders model the channel corruption as a forward stochastic process and learn to reverse it via a sequence of denoising steps, achieving impressive gains under various noise conditions \citep{choukroun2022denoisingdiffusionerrorcorrection}.  Meanwhile, Transformer-based decoders exploit self-attention to capture long-range code constraints; notable examples include the Error Correction Code Transformer with its extensions  \citep{choukroun2022errorcorrectioncodetransformer, choukroun2024deep, choukroun2024foundation, choukroun2024learning} and recent variants employing layer-wise masking and cross‐message‐passing modules to enhance both expressivity and decoding speed \citep{park2023mask,park2024crossmpt}.

\subsection{Mamba Architecture}
In recent years, State‐Space Models (SSMs) have attracted considerable attention as an alternative to purely attention‐based architectures for sequence modeling, due to their ability to capture long‐range dependencies with favorable computational and memory efficiency \citep{gu2021efficiently,gu2021combining}.  A landmark contribution in this domain is the Structured State Space Sequence (S4) model, which leverages parameterized linear dynamical systems and the HiPPO framework \citep{gu2020hippo} to achieve expressive, convolutional representations of sequential data.  Building upon S4, subsequent work proposed the Mamba architecture, wherein the SSM’s convolutional kernels are dynamically generated as functions of the input sequence \citep{gu2023mamba,dao2024transformers}.  Empirical evaluations demonstrate that Mamba attains inference speeds up to five times faster than comparable Transformer models while scaling seamlessly to input lengths on the order of millions of elements. Moreover, when Mamba is integrated with Transformer layers in a hybrid configuration, the resulting model consistently surpasses both standalone Transformer and S4 architectures in a range of language and time‐series benchmarks.

\section{Background}
In this section, we formalize the decoding setup for binary linear block codes using the notation of \cite{choukroun2022errorcorrectioncodetransformer}.  Let \(C\subseteq \mathbb{F}_2^n\) be a binary linear block code of length \(n\) and dimension \(k\), defined by its parity‐check matrix $H \in \mathbb{F}_2^{(n-k)\times n}$. A vector \(x\in\mathbb{F}_2^n\) is a valid codeword if and only if $ H\,x \;=\; 0\,$.
Transmission occurs over an Additive White Gaussian Noise (AWGN) channel with Binary Phase‐Shift Keying (BPSK) modulation.  Under this model, the codeword \(x\in\{0,1\}^n\) is mapped to \(x_s\in\{\pm1\}^n\subset\mathbb{R}^n\) and corrupted by Gaussian noise \(z\sim\mathcal{N}(0,\sigma^2 I_n)\), yielding the received vector $y \;=\; x_s \;+\; z\,$. To enforce invariance to the transmitted codeword and mitigate overfitting, we construct the decoder input from the magnitude of the channel output and its syndrome as in \citep{bennatan2018deep}.  First, we obtain the hard‐decision vector $y_b \;=\; \tfrac{1-\operatorname{sign}(y)}{2}\;\in\;\{0,1\}^n,$ and then compute the syndrome $s \;=\; H\,y_b \;\in\;\mathbb{F}_2^{\,n-k}.$ Finally, we concatenate the amplitude \(|y|\in\mathbb{R}^n\) with the syndrome \(s\) to form the decoder input $y_{\mathrm{in}}
  \;=\;\bigl[\,|y|;\,s\,\bigr]
  \;\in\;\mathbb{R}^{\,n+(n-k)}\,,$ which is provided to the proposed Mamba–Transformer decoder.  

\section{Method - Mamba-Transformer Decoder}

\begin{figure}[htbp]
\vspace{0pt}

\begin{minipage}[t]{0.24\linewidth}%
\vspace{30pt}
\includegraphics[width=\linewidth]{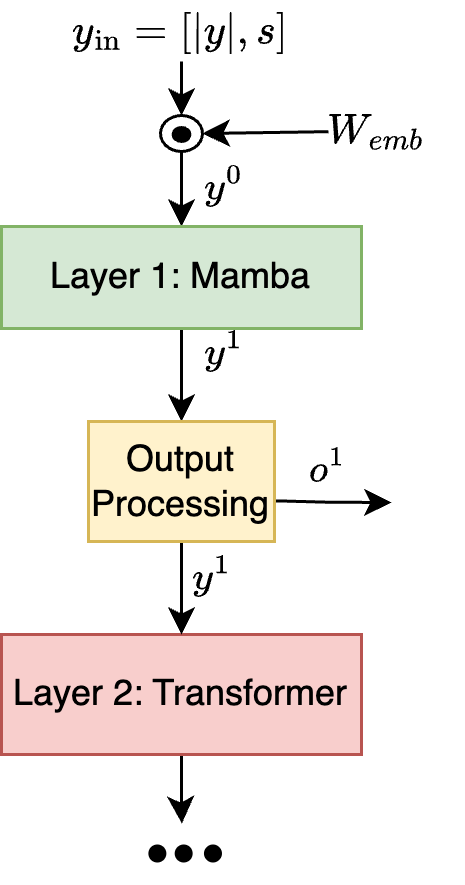}%
\subcaption{Model Blocks}
\end{minipage}
\begin{minipage}[t]{0.75\linewidth}
\vspace{0pt}
    \begin{minipage}[t]{\linewidth}
        \vspace{0pt}
        \includegraphics[width=\linewidth]{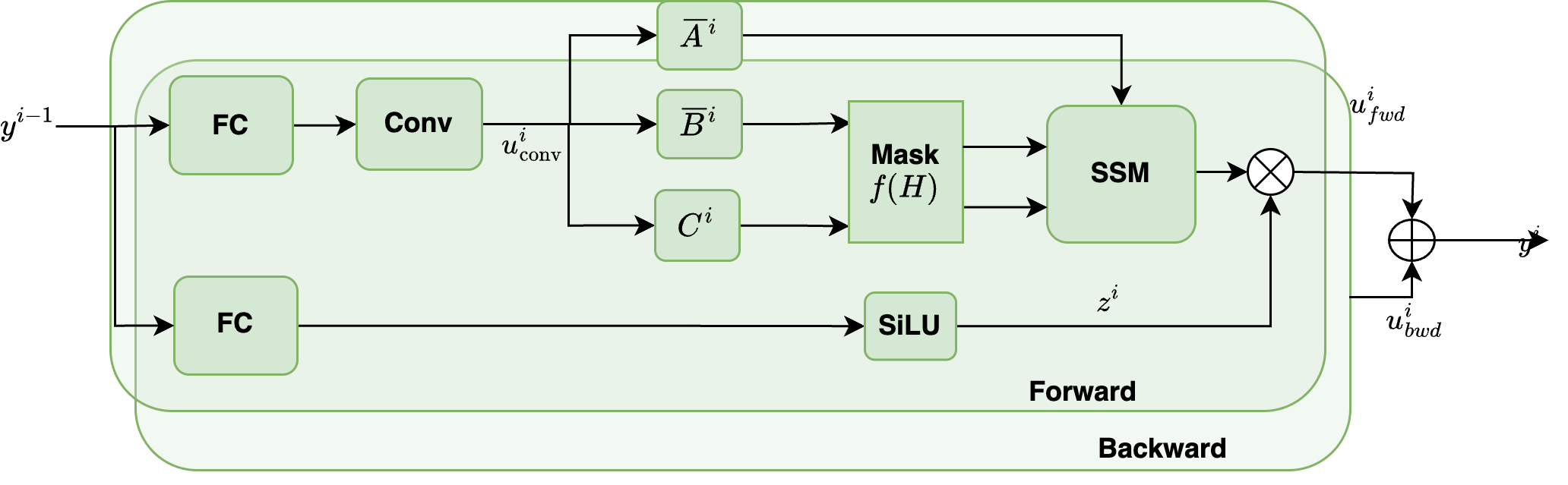}
        \subcaption{Mamba block structure}
    \end{minipage}
    
    \begin{minipage}[t]{\linewidth}
        \includegraphics[width=\linewidth]{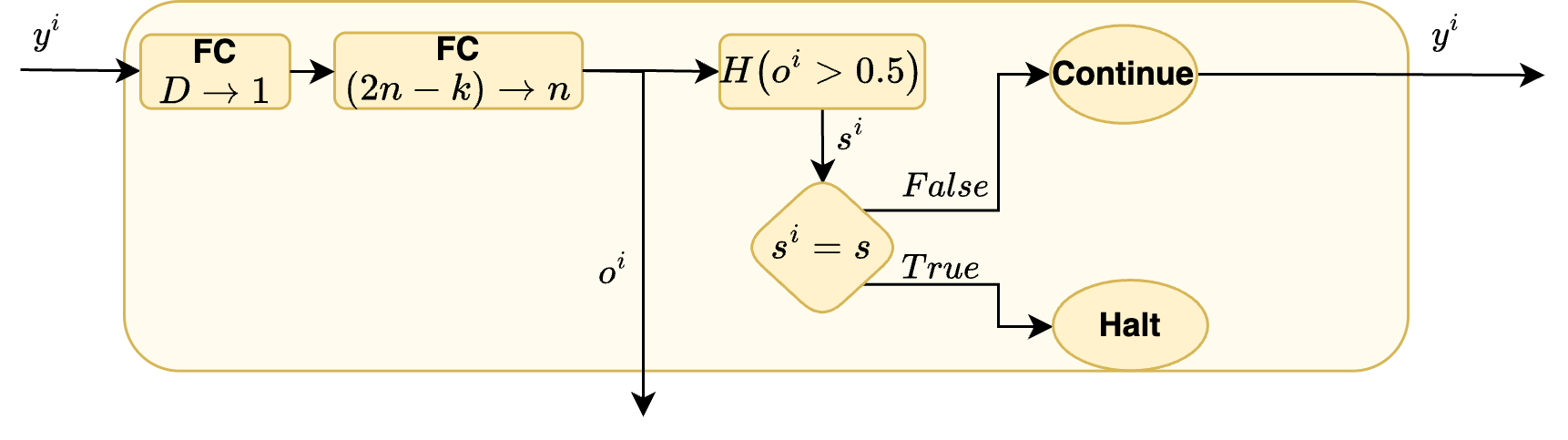}
        \subcaption{Model output processing}
    \end{minipage}
    \begin{minipage}[t]{\linewidth}
        \includegraphics[width=\linewidth]{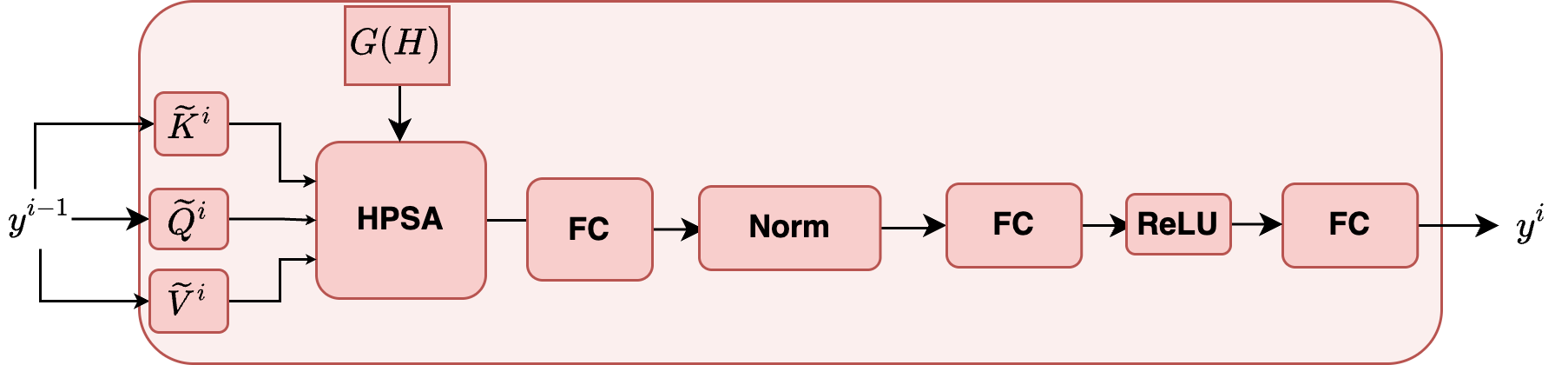}
        \subcaption{Attention block structure}
    \end{minipage}

\end{minipage}

\centering
\caption{EECM architecture}
\end{figure}

The ECCM model is composed from $N_{layers}$ layers, the layers are alternating between mamba layers and attention layers, starting with a mamba layer. The layers $l_i$ where $i\in{1,3,...}$ are Mamba layers and the layers $l_i$ where $i \in {2,4,...}$ are attention blocks.
The input to the $l_i$ layer is denoted $y^{i-1}$, with 
\begin{equation}
\label{eq:init_y0}
    y^0=y_{in} \odot W_{emb}
\end{equation}
where $W^{i}_{emb}\in \mathbb{R}^{L \times D}$, where $D$ is the model hidden dimension, $L=2n-k$, and $\odot$ represents the element-wise multiplication operator.  Note that $y^i$ for $i > 0$, will be the output of the $i$-th layer.
The architecture uses two masks that are produced from the parity check matrix, $f(H) \in \mathbb{Z}_2^{(n-k)\times(2n-k)}$ and $g(H)\in \mathbb{Z}_2^{(2n-k)\times(2n-k)}$ which are used by the mamba layers and attention layers - respectively. The $g(H)$ mask is taken from \citep{choukroun2022errorcorrectioncodetransformer}:
\begin{gather}
    g(H) = 
    \begin{bmatrix}
    Graph(H) & H^T \\
    H & I_{n-k}
    \end{bmatrix}
\end{gather}
and $Graph(H) \in \mathbb{R}^{n \times n}$
\begin{equation}
\mathrm{Graph}(H)[i, j] =
\begin{cases}
1, & \exists\, m \in [1, n - k] \text{ such that } H[m, i] = 1 \text{ and } H[m, j] = 1 \\
0, & \text{otherwise}
\end{cases}
\end{equation}

The $g(H)$ mask, ensures that only pairs that are on the same parity check line are computed, which reduces the complexity of the attention operation and induces knowledge of the code into the model.

The proposed $f(H)$ mask, is devised to have the same effect on the SSM process. By applying it to the matrices of the operation, the effect of the input in a specific position only changes the state for bits that are on the same parity check line. Ensuring that interactions only happens along the parity check lines  
\begin{equation}
    f(H) = [H;I_{n-k}]
\end{equation}

\subsubsection{Mamba block}
Each Mamba block contains the following operations:
First  $y^{i-1}$ is projected with two learnable matrices $W^{i}_u ,W^{i}_z \in \mathbb{R}^{D \times D}$:
\begin{equation}
\label{eq:mamba:start}
    u^{i} = W_u^{i} y^{i-1}
\end{equation}
\begin{equation}
\label{eq:zgen}
    z^{i} = SiLU(W_z ^ {i} y^{i-1})
\end{equation}
where $u^{i},z^{i} \in \mathbb{R}^{L\times D}$. 
and SiLU  is the activation function \citep{hendrycks2023gaussianerrorlinearunits}. Then apply 1D-Convolution layer $Conv^{i}$ to $u^{i}$ over the sequence length:
\begin{equation}
    u^{i}_{conv} = {Conv}^{i} \left(  {u^{i}}\right)
\end{equation}
where $u^{i}_{conv} \in \mathbb{R}^{L\times D}$. Then we apply the Selective-State-Space Model \cite{gu2023mamba} with modification to the error-correcting code scenario. First, $u^{i}_{conv}$ projected to $B^{i},C^{i} \in \mathbb{R}^{L\times S}$ where $S$ is the dimension of the state.
\begin{equation}
    B^{i}= u^{i}_{conv}{W^{i}_b}^T
\end{equation}
\begin{equation}
    C^{i}=u^{i}_{conv}{W^{i}_c}^T
\end{equation}
Where $W^{i}_b,W^{i}_c \in \mathbb{R}^{S \times D}$ are learnable matrices. Then we apply the discretization process (as described in \cite{gu2023mamba}) on matrices $B^i$, and $A^i$ where $A^i \in \mathbb{R}^{D \times S}$ which is a learnable matrix. First, generate the $\Delta^i \in \mathbb{R}^{L \times D}$ matrix:
\begin{equation}
    \Delta^i = W_{\Delta}^{i}u^{i}_{conv}
\end{equation}

where $W_{\Delta}^{i} \in \mathbb{R}^{L \times D}$ is a learnable matrix. 
Second, initialize the tensors $\bar{A}^{i},\bar{B}^{i}\in \mathbb{R}^{L \times D \times S }$, 
\begin{gather}
 \bar{A}^{i}[l,s,d] = exp(A^i[l,s]\Delta^i[l,d])  \\
 \bar{B}^{i}[l,s,d] = B^i[l,s]\Delta^i[l,d]
\end{gather}
 
where $l\in[1,L],s\in[1,S],d\in[1,D]$. Here the error-code specific modification is inserted, using a mask. Generate the mask matrix $f(H) = [H;I_K]$ where $I_K\in\mathbb{R}^{K\times K}$ is the identity matrix. Then apply the mask to the matrices $\bar{B}^{i}$ and $C^{i}$ creating the matrices $\bar{B}^{i}_{M}$ and $C^{i}_{M}$, respectively.

\begin{equation}
    \bar{B}^{i}_{M}[l,d,s] = \begin{cases}
        f(H)[l,d]\bar{B}^{i}[l,d,s]&  d < (n-k)\\
        0&  otherwise
    \end{cases}
\end{equation}

where $l\in[1,L],s\in[1,S],d\in[1,D]$
\begin{equation}
    {C}^{i}_{M}[l,s] = \begin{cases}
        f(H)[l,s]{C}^{i}[l,s]&  s < (n-k)\\
        0&  otherwise
    \end{cases}
\end{equation}

Apply the SSM process in which a series of states $h_{l} \in \mathbb{R}^{D\times S}$ are calculated:

\begin{equation}
    h_{l}[d,s] = \bar{A}[l,d,s] h_{l-1}[d,s] + \bar{B}_M^{i}[l,d,s]u^{i}_{conv}[l,d] 
\end{equation}
\begin{equation*}
        u^{i}_{ssm}[l,d] = \sum_{i=1}^{S}{h_{l}[d,i]C[l,i]} + R[d]u^{i}_{conv}[l,d]
\end{equation*}
where $R\in\mathbb{R}^F$ is a learnable vector, And initialize $h_0$ to a vector of zeros.

Then we apply the gating from Eq.\ref{eq:zgen}:
\begin{equation}
\label{eq:mamba:gate}
    u^{i}_{fwd}= z^{i} \odot u^{i}_{ssm}
\end{equation}

Up to this point, the description was of the processing in the causal direction.
Now apply the reverse direction processing in order to achieve a Bi-Directional Mamba.
Meaning, substitute $y^{i-1}$,$H$ and $u^{i}_{fwd}$ to
 $\overleftarrow{y}^{i-1}$, $\overleftarrow{f(H)}$ and $\overleftarrow{u}^{i}_{fwd}$.
And apply Eq.\ref{eq:mamba:start} through Eq.\ref{eq:mamba:gate}
where:
\begin{equation}
    \overleftarrow{y}^{i-1}[l,d] = y^{i-1}[L-l,d]
\end{equation}
\begin{equation}
    \overleftarrow{f(H)}[l,d] =f(H)[L-l,d] 
\end{equation}
and $\overleftarrow{u}^{i}_{fwd}$ is the output of the process.
Then calculate $u^{i}_{bwd} \in \mathbb{R}^{L\times D}$
\begin{equation}
    u^{i}_{bwd}[l,d] = \overleftarrow{u}^{i}_{fwd}[L-l,d]
\end{equation}
and then the output of the block $y^{i}\in \mathbb{R}^{L\times D}$ is calculated.
\begin{equation}
    y^{i} = u^{i}_{fwd} + u^{i}_{bwd}
\end{equation}

\subsection{Transformer Block}
First compute the queries, keys, and values $Q^i,K^i,V^i \in \mathbb{R}^{L \times D}$
using the the input $y^{i-1}$:
\begin{equation}
\begin{aligned}
Q^i &= y^{i-1} W_Q^i + b_Q^i \\
K^i &= y^{i-1} W_K^i + b_K^i\\
V^i &= y^{i-1} W_V^i + b_V^i
\end{aligned}
\end{equation}
where $W_Q^i, W_K^i, W_V^i \in \mathbb{R}^{D\times D}$ and $b_Q^i, b_K^i, b_V^i \in \mathbb{R}^{D}$ are learned parameters. These are reshaped into $h$ attention heads with per-head dimension $d_k = D / h$:
\begin{equation}
\tilde{Q}^i, \tilde{K}^i, \tilde{V}^i \in \mathbb{R}^{h \times L \times d_k}.
\end{equation}
Then apply the HPSA mechanism as described in \citep{levy2025accelerating}:
\begin{equation}
\tilde{O}^i, \alpha^i = \mathrm{HPSA}(\tilde{Q}^i, \tilde{K}^i, \tilde{V}^i, \text{g(H)}),
\end{equation}
with \( \tilde{O}^i \in \mathbb{R}^{h \times L \times d_k} \) and \( \alpha^i \in \mathbb{R}^{h \times L \times L} \). The outputs from all heads are concatenated $O^i  \in \mathbb{R}^{L \times D} $:
\begin{equation}
O^i = \mathrm{concat}(\tilde{O}^i).
\end{equation}
Finally, the output of the attention block is computed as $y^i_{a} \in \mathbb{R}^{L \times D}$:
\begin{equation}
y_a^{i} = O^i W_O^i + b_O^i
\end{equation}
where $W_O^i \in \mathbb{R}^{D \times D}$ and $b_O^i \in \mathbb{R}^{D}$. Then apply layer norm \citep{ba2016layernormalization} to calculate $\tilde{y}^i_a \in \mathbb{R}^{L \times D}$
\begin{equation}
    \tilde{y}^i_a = LayerNorm(y_a^i)
\end{equation}
then $y^i  \in \mathbb{R}^{L \times D}$ is calculated:
\begin{equation}
    y^i = ReLU(\tilde{y}^i_aW^i_1 + b_1)W_2 + b_2
\end{equation}
where $W_1 \in \mathbb{R}^{D \times 4D},\ b_1 \in \mathbb{R}^{4D},\ W_2 \in \mathbb{R}^{4D \times D},\ b_2 \in \mathbb{R}^D$ are learnable parameters.

\subsection{Model Output}

After each layer, $y^{i}$ is project down to $o^{i} \in \mathbb{R}^{N}$ using $w_r \in \mathbb{R}^D$,$b_r\in\mathbb{R}^L$,$W_s \in \mathbb{R}^{N \times L}$,$b_s \in \mathbb{R}^N$ which are learnable parameters:
\begin{equation}
    o^{i} = \sigma(W_s(y^{i}w_r +b_r) +b_s)
\end{equation}
where $\sigma$ is the sigmoid function. The syndrome is calculated:
\begin{equation}
\label{eq:early_stopping}
    s^{i} = H(o^{i} > 0.5)
\end{equation}
if $s^{i} = s$ the processing is stopped - and set $i_{last} = i$,
if $s^{i}\ \ne\ s\ \forall\ i\ \in\ [1,N_{layers}]$ set $i_{last} = N_{layers}$

Note that the model output is an estimate for the input's multiplicative noise,
therefore in order to calculate the estimated code-word:
\begin{equation}
    \hat{c}^{i}[l] = \frac{1 - sign((1 - 2 o^i[l]) y_{in}[l])}{2}    
\end{equation}

\subsection{Loss Function}
In order to calculate the loss, first calculate in which bit an error occured as in \cite{bennatan2018deep}
\begin{equation}
    z[l] = \frac{1 - sign((1 - 2 c[l]) x_{in}[l])}{2}
\end{equation}

Then calculate the Binary Cross Entropy (BCE) between $o^i$ and $z$, and sum over all the outputs.
\begin{equation}
    Loss=\sum_i^{N_{layers}} BCE(o^{i},z)
\end{equation}
Note that if for some $i_0<N_{blocks}$, $Ho^{i_0} = 0$, the outputs for $i>i_0$ are not calculated and therefore not summed in the loss.

\section{Experiments}

To evaluate the proposed decoder, we train it on four classes of linear block codes: Bose–Chaudhuri–Hocquenghem (BCH) codes~\citep{bose1960class}, Low-Density Parity-Check (LDPC) codes~\citep{gallager2003low}, Polar codes~\citep{arikan2009channel}, and MacKay codes. The corresponding parity-check matrices are obtained from~\cite{ParityCheckMatrix}. Training data are generated at six signal-to-noise ratio (SNR) levels, $\mathrm{SNR}\in\{2,\dots,7\}$ dB, then added to the generated message to simulate a AWGN channel. We use the zero-codeword in the training process in order to verify that the model doesn't overfit the codewords it sees, by simply changing to random codewords on model evaluation. The Adam \citep{kingma2017adammethodstochasticoptimization} optimizer was configured with a learning rate of $2.5\times10^{-4}$ and decays to $10^{-10}$ following a cosine \citep{loshchilov2017sgdrstochasticgradientdescent} schedule. The training was done with batch size of 128 and 1000 batches per epoch. 
In all the experiments we set $D=128, N_{blocks}=8,h=8, S=128$, where $D$ is the embedding size, $S$ is the Mamba block's state size, $h$ is the number of attention heads, and $N_{blocks}$ is the number of blocks, meaning there are 4 Mamba blocks and 4 attention blocks, the resulting model has similar number of parameters to previous methods at approx $1.2M$. 
For evaluation, we simulate test examples at SNR levels of 4dB, 5dB, and 6dB, and report the negative natural logarithm of the bit error rate, $-\ln\bigl(\mathrm{BER}\bigr).$ Each evaluation run is continued until a fixed number of decoding errors, $500$, has been observed similar to \citep{park2024crossmpt}.

\section{Results and Discussions}

\begin{figure}
    \centering
    \begin{subfigure}[b]{0.48\linewidth}
        \includegraphics[width=\linewidth]{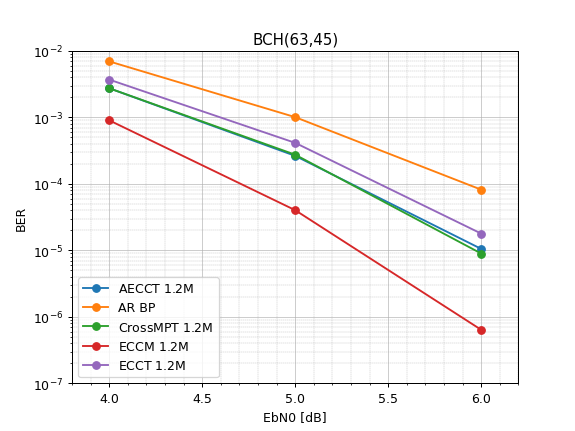}
        \subcaption{BCH Code N=63 K=45}
    \end{subfigure}
    \begin{subfigure}[b]{0.48\linewidth}
        \includegraphics[width=\linewidth]{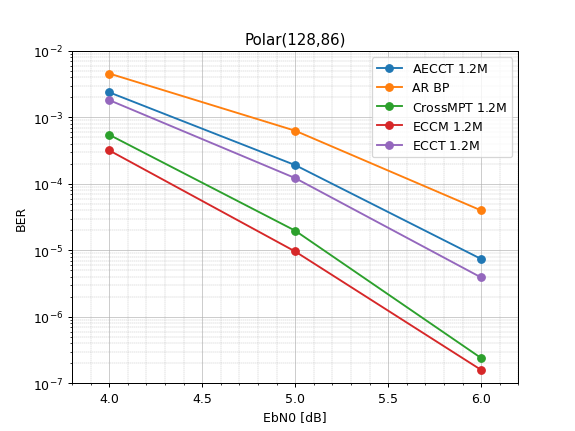}
        \subcaption{Polar Code N=128 K=86}
    \end{subfigure}
    \caption{BER-SNR performance of ECCM versus baselines, on BCH and POLAR codes}
    \label{fig:ber-snr}
\end{figure}

In Tab.\ref{tab:results}, the results are the presented compared to previous methods, for each code 6 methods are presented: BP, ARBP, ECCT, AECCT, CrossMPT, and our own method ECCM. For each, the table shows the negative natural log of the BER at SNR levels 4dB, 5dB, and 6dB. The best method is marked in \textbf{bold}, in places where reported results were not available the "-" mark was used. The table shows that ECCM consistently outperforms all the other methods across all BCH codes, and SNR levels. Notably outperforming CrossMPT - with a significant improvement in the decoding of BCH(63,45) code, achieving over $18\%$ in terms of negative natural logarithm of BER, $-ln(\text{BER})$, ECCM shows comparable performance to CrossMPT in the decoding of the Polar(64,48) code, and shows notable improvements in longer Polar codes - achieveing up to $7.2\%$ gain in the Polar(128,86) code. While CrossMPT achieves better results in some of the LDPC codes the improvement are modest typically around $4\%$, ECCM achieves better performance in decoding LDPC(49,24) and - comparable to increased - performance on  LDPC(121,80). It also outperforms all other models in the MacKay Code, slightly outperforming CrossMPT, which indicates the model is capable in learning very sparse parity-check matrices. Fig \ref{fig:ber-snr} shows the performence in terms of BER as a function of SNR for the above methods. It is important to note that integrating ECCM and CrossMPT is possible, which may close the gap in LDPC codes decoding.

\begin{table*}[!t]
\caption{Comparison of decoding performance at three SNR values (4, 5, 6) for BP, ARBP~\citep{nachmani2021autoregressive}, ECCT~\citep{choukroun2022errorcorrectioncodetransformer}, AECCT~\citep{levy2025accelerating}, CrossMPT~\citep{park2024crossmpt}, and ECCM. The results are measured by the negative natural logarithm of BER ($-\ln(\mathrm{BER})$). The best results are highlighted in \textbf{bold}. Higher is better.}
\label{tab:results}
\begin{center}
\begin{small}
\resizebox{\textwidth}{!}{
\begin{tabular}{cccccccccccccccccccc}
\toprule
\multirow{2}{*}{\textbf{Codes}} & \multirow{2}{*}{\textbf{$(N,K)$}}
  & \multicolumn{3}{c}{\textbf{BP}}
  & \multicolumn{3}{c}{\textbf{AR\,BP} ~\citep{nachmani2021autoregressive}}
  & \multicolumn{3}{c}{\textbf{ECCT $1.2M$} ~\citep{choukroun2022errorcorrectioncodetransformer}}
  & \multicolumn{3}{c}{\textbf{AECCT  $1.2M$}~\citep{levy2025accelerating}}
  & \multicolumn{3}{c}{\textbf{CrossMPT  $1.2M$}~\citep{park2024crossmpt}}
  & \multicolumn{3}{c}{\textbf{ECCM  $1.2M$} (ours)} \\
\cmidrule(lr){3-5}\cmidrule(lr){6-8}\cmidrule(lr){9-11}\cmidrule(lr){12-14}\cmidrule(lr){15-17}\cmidrule(lr){18-20}
 & & 4 & 5 & 6
   & 4 & 5 & 6
   & 4 & 5 & 6
   & 4 & 5 & 6
   & 4 & 5 & 6
   & 4 & 5 & 6 \\
\midrule
\multirow{4}{*}{BCH}
 & (31,16)  & \makecell{4.63\\–}    & \makecell{5.88\\–}    & \makecell{7.60\\–}
            & 5.48 & 7.37 & 9.60
            & 6.39 & 8.29 & 10.66
            & 7.01 & 9.33 & 12.27
            & 6.98 & 9.25 & 12.48
            & \textbf{7.26} & \textbf{9.71} & \textbf{12.66} \\
 & (63,36)  & \makecell{3.72\\4.03} & \makecell{4.65\\5.42} & \makecell{5.66\\7.26}
            & 4.57 & 6.39 & 8.92
            & 4.68 & 6.65 & 9.10
            & 5.19 & 6.95 & 9.33
            & 5.03 & 6.91 & 9.37
            & \textbf{5.49} & \textbf{7.52} & \textbf{10.23} \\
 & (63,45)  & \makecell{4.08\\4.36} & \makecell{4.96\\5.55} & \makecell{6.07\\7.26}
            & 4.97 & 6.90 & 9.41
            & 5.60 & 7.79 & 10.93
            & 5.90 & 8.24 & 11.46
            & 5.90 & 8.20 & 11.62
            & \textbf{7.01} & \textbf{10.12} & \textbf{14.26} \\
 & (63,51)  & \makecell{4.34\\4.50} & \makecell{5.29\\5.82} & \makecell{6.35\\7.42}
            & 5.17 & 7.16 & 9.53
            & 5.66 & 7.89 & 11.01
            & 5.72 & 8.01 & 11.24
            & 5.78 & 8.08 & 11.41
            & \textbf{6.10} & \textbf{8.77} & \textbf{12.22} \\
\midrule
\multirow{3}{*}{Polar}
 & (64,48)  & \makecell{3.52\\4.26} & \makecell{4.04\\5.38} & \makecell{4.48\\6.50}
            & 5.41 & 7.19 & 9.30
            & 6.36 & 8.46 & 11.09
            & 6.43 & 8.54 & 11.12
            & 6.51 & \textbf{8.70} & \textbf{11.31}
            & \textbf{6.61} & 8.61 & 11.20 \\
 & (128,86) & \makecell{3.80\\4.49} & \makecell{4.19\\5.65} & \makecell{4.62\\6.97}
            & 5.39 & 7.37 & 10.13
            & 6.31 & 9.01 & 12.45
            & 6.04 & 8.56 & 11.81
            & 7.51 & 10.83 & 15.24
            & \textbf{8.05} & \textbf{11.55} & \textbf{15.65} \\
 & (128,96) & \makecell{3.99\\4.61} & \makecell{4.41\\5.79} & \makecell{4.78\\7.08}
            & 5.27 & 7.44 & 10.20
            & 6.31 & 9.12 & 12.47
            & 6.11 & 8.81 & 12.15
            & 7.15 & 10.15 & 13.13
            & \textbf{7.49} & \textbf{10.45} & \textbf{13.27} \\
\midrule
\multirow{3}{*}{LDPC}
 & (49,24)  & \makecell{5.30\\6.23} & \makecell{7.28\\8.19} & \makecell{9.88\\11.72}
            & 6.58 & 9.39 & 12.39
            & 5.79 & 8.13 & 11.40
            & 6.10 & 8.65 & 12.34
            & 6.68 & 9.52 & 13.19
            & \textbf{6.71} & \textbf{9.55} & \textbf{13.25} \\
 & (121,60) & \makecell{4.82\\–}    & \makecell{7.21\\–}    & \makecell{10.87\\–}
            & 5.22 & 8.31 & 13.07
            & 5.01 & 7.99 & 12.78
            & 5.17 & 8.32 & 13.40
            & \textbf{5.74} & \textbf{9.26} & \textbf{14.78}
            & 5.49 & 8.87 & 14.23 \\
 & (121,80) & \makecell{6.66\\–}    & \makecell{9.82\\–}    & \makecell{13.98\\–}
            & 7.22 & 11.03 & 15.90
            & –    & –     & –
            & –    & –     & –
            & \textbf{7.99} & \textbf{12.75} & 18.15
            & 7.81 & 12.34 & \textbf{18.35} \\
\midrule
MacKay       & (96,48)   & \makecell{–\\–}      & \makecell{–\\–}      & \makecell{–\\–}
             & 7.43 & 10.65 & 14.65
             & –    & –     & –
             & –    & –     & –
             & 7.97 & 11.77 & 15.52
             & \textbf{7.98} & \textbf{11.84} & \textbf{15.70} \\
\bottomrule
\end{tabular}
}
\end{small}
\end{center}
\end{table*}


\section{Model Analysis}

\subsection{Ablation Analysis}
To analyze the contribution of each of the following proposed modifications: combining Mamba and Transformer, using loss from every layer, and the proposed mask for the mamba layers, variants of the proposed method were trained - removing one modification at a time. The variants were all trained on the BCH(63,45) code, with the same hyper-parameters as discussed above. The same performance evaluation process was used in this study as well. Table \ref{tab:ablation} demonstrates that each of the proposed modifications contributes to the performance of the final model. The top row shows the full method, and in each subsequent row, one modification is removed to isolate its effect. The "Mamba Mask" indicates which mask was used in the experiment in the mamba layers, either the proposed mask $f(H)$, or the baseline mask $g(H)$ from ECCT \citep{choukroun2022errorcorrectioncodetransformer}. In experiment (iii) no Mamba layers are used. The "Model Layout" column indicates whether in the experiment Mamba and Transformer layers were used, or only Transformer layers. The "Multi-loss" column is "True" if the loss was computed using the output from each layer, and "False" where it was computed only on the output of the last layer. Experiment (i) shows that using the proposed mask $f(H)$ yields better results than using $g(H)$. Experiment (ii) demonstrates that using the loss from each layer is contributing significantly to the proposed model's performance. Experiment (iii)  shows that combining the mamba and transformer layers yield better results, compared to both experiment (i) and the proposed model, confirming that it the modification is an improvement regardless of the used mask.

\begin{table}[H]
\centering
\caption{Ablation analysis: negative natural logarithm of bit error rate (BER) for our complete method compared with its partial components. Higher values indicate better performance, Highest value is marked in \textbf{bold}. 
}
\label{tab:ablation}
\renewcommand{\arraystretch}{1.2}
\begin{tabular}{@{} lllc *{3}{c} @{}}
\toprule
\multirow{2}{*}{Experiment} & \multirow{2}{*}{Mamba Mask} & \multirow{2}{*}{Model Layout} & \multirow{2}{*}{Multi-Loss} & \multicolumn{3}{c}{SNR (dB)} \\ 
\cmidrule(l){5-7}
 &  &  &  & 4 & 5 & 6 \\
\midrule
\textbf{Full Method} & $f(h)$     & Transformer \& Mamba & True  & \textbf{7.01}  & \textbf{10.12} & \textbf{14.26} \\
\midrule
(i)   & $g(H)$     & Transformer \& Mamba & True  & 6.86  & 9.88  & 13.76 \\
(ii)  & $f(H)$     & Transformer \& Mamba & False & 5.80  & 8.18  & 11.60 \\
(iii) & N/A        & Transformer only     & True  & 6.66  & 9.45  & 13.31 \\
\bottomrule
\end{tabular}
\end{table}

\subsection{Complexity Analysis}
The complexity of the proposed method can be separated into the complexity of the the transformer block and the complexity of the Mamba block. The complexity of the transformer block is $O((2n-k)D^2 + (2n-k)^2D\rho(G(H)))$ where $\rho(H)$ is the sparsity of the mask matrix. Moreover, complexity of the Mamba blocks is $O((2n-k)DS)$, the total complexity of the model is $O(N_{blocks}(2n-k)D(N_{Mamba}S + N_{transformer}(D+(2n-k)\rho(G(H))))$, since the $N_{Mamba} =N_{transformer} = \frac{1}{2} N_{blocks}$, we expect a significant speedup relative to AECCT and CrossMPT which is $O((2n-k)D^2+n(n-k)(\rho(H))$ \citep{park2024crossmpt}.
\begin{table}[H]
\centering
\caption{Average inference speed in $\mu s$ for ECCT, AECCT, ECCM, and CrossMPT. Lower is better.}
\label{tab:inference-speed:model-compare}
\setlength{\tabcolsep}{8pt}
\renewcommand{\arraystretch}{1.2}
\begin{tabular}{@{} l r r r r @{}}
\toprule
\textbf{Code} & \multicolumn{1}{c}{\textbf{ECCT}} & \multicolumn{1}{c}{\textbf{AECCT}} & \multicolumn{1}{c}{\textbf{CrossMPT}} & \multicolumn{1}{c}{\textbf{ECCM}} \\
\toprule
LDPC (49,24) & 237.53 $\mu s$ & 229.21 $\mu s$ & 222.57 $\mu s$ & \textbf{72.08} $\mu s$ \\
LDPC (121,60) & 332.23 $\mu s$ & 358.81 $\mu s$ & 289.11 $\mu s$ & \textbf{260.42} $\mu s$ \\
BCH (31,16) & 222.57 $\mu s$ & 232.56 $\mu s$ & 217.25 $\mu s$ & \textbf{58.78} $\mu s$ \\
BCH (63,45) & 239.18 $\mu s$ & 251.39 $\mu s$ & 227.28 $\mu s$ &\textbf{75.42} $\mu s$ \\
Polar (64,48) & 239.18 $\mu s$ & 249.63 $\mu s$ & 227.48 $\mu s$ & \textbf{72.34} $\mu s$ \\
Polar (128,96) & 315.56 $\mu s$ & 332.23 $\mu s$ & 272.41 $\mu s$ & \textbf{221.98} $\mu s$ \\
\bottomrule
\end{tabular}
\end{table}

Validating the claim that the suggested method is more efficient, the following experiment was performed. For each of the methods ECCT, AECCT, CrossMPT, and ECCM - on the same machine apply inference on batches of codewords, with the same number of batches, measure the time that the process took and divide by the number of codewords generated. The process was performed with a batch size of 512 and 200 batches, on a machine with a NVIDIA GeForce RTX 4090 GPU and a Intel i9-14400K. For this test the Early Stopping feature of the model was disabled.

\begin{table}[H]
\centering
\caption{Inference speed at diferent SNRs. Lower is better, the best time is marked in bold.}
\label{tab:inference-speed:early-stopping}
\setlength{\tabcolsep}{8pt}
\renewcommand{\arraystretch}{1.2}
\begin{tabular}{@{} l r r r *{4}{r} @{}}
\toprule
\multirow{1}{*}{\textbf{Code}} 
  & \multicolumn{1}{c}{\textbf{No Early Stopping}} 
  & \multicolumn{1}{c}{\textbf{4 [dB]}} 
  & \multicolumn{1}{c}{\textbf{5 [dB]}}
  & \multicolumn{1}{c}{\textbf{6 [dB]}} \\
\toprule
LDPC (49,24) & {72.08} $\mu s$ & {70.26} $\mu s$ & {60.85} $\mu s$ & {59.74} $\mu s$ \\
LDPC (121,60) & {260.42} $\mu s$ & {260.42} $\mu s$ & {253.68} $\mu s$ & {145.78} $\mu s$ \\
BCH (31,16) & {58.78} $\mu s$ & {57.79} $\mu s$ & {57.79} $\mu s$ & {57.79} $\mu s$ \\
BCH (63,45) & {75.42} $\mu s$ & {72.89} $\mu s$ & {62.61} $\mu s$ & {62.61} $\mu s$ \\
Polar (64,48) & {72.34} $\mu s$ & {70.01} $\mu s$ & {62.31} $\mu s$ & {61.62} $\mu s$ \\
Polar (128,96) & {221.98} $\mu s$ & {219.5} $\mu s$ & {197.32} $\mu s$ & {119.11} $\mu s$ \\
\bottomrule
\end{tabular}
\end{table}

On the same machine, a similar test was performed but instead of testing on the different methods, the test was performed only on ECCM and at different SNRs [dB]: 4, 5, 6. The intention of this test is to show that the early-stopping feature is meaningful for inference runtime. The test results (Table \ref{tab:inference-speed:early-stopping} indicates that a partial model can still be used without major performance degradation for SNRs 5[dB] and 6[dB], since the speedup in these cases is coming from layers of the model not being used.

\subsection{Attention Score Comparison}
\begin{wraptable}{r}{0.57\textwidth}
  \centering
  \begin{subfigure}[b]{0.49\linewidth}
    \centering
    \includegraphics[width=\linewidth]{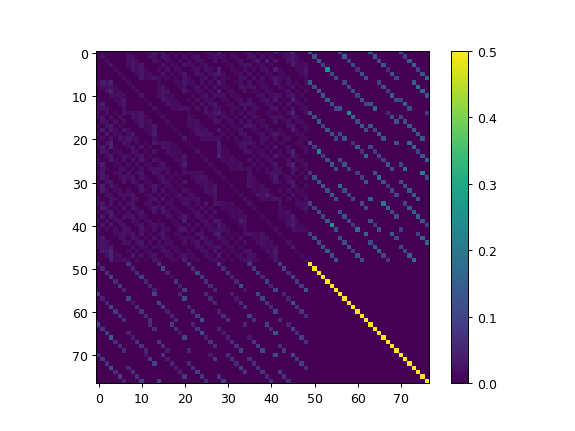}
    \caption{No error}
    \label{fig:attention_map_no_error}
  \end{subfigure}
  \hfill
  \begin{subfigure}[b]{0.49\linewidth}
    \centering
    \includegraphics[width=\linewidth]{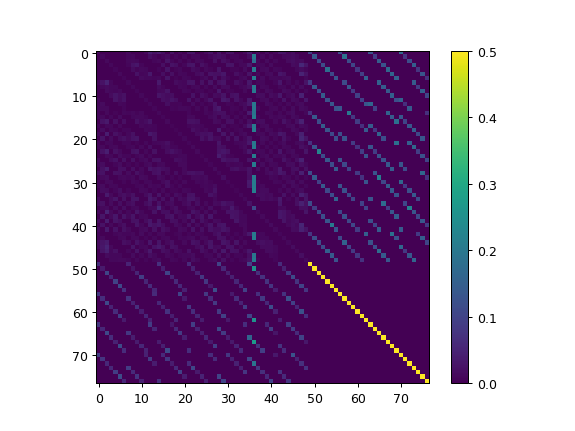}
    \caption{Single error}
    \label{fig:attention_map_single_error}
  \end{subfigure}
  \caption{Comparison of attention maps sum. (a) without any error (b) with a single error. }
  \label{fig:attention_map_comparison}
\end{wraptable}
In order to compare our model's behavior with ECCT \citep{choukroun2022errorcorrectioncodetransformer}, we will examine the internal attention scores of the model's layers in two cases, one where there is no error in the input, and the other where there is a single error in the input. Using this method will show how the attention reacts to error in the input. To visualize the attention across the model, compute the full forward pass of the model with the two inputs, and sum the attention scores across the transformer blocks of the model. For this experiment, evaluate all the layers regardless if the syndrome condition is met in Eq. \ref{eq:early_stopping}. Examining the attention maps Fig \ref{fig:attention_map_comparison}, we can identify four distinct regions corresponding to the structure of the $g(H)$ mask: \emph{magnitude $\rightarrow$ magnitude} (top-left), \emph{magnitude $\rightarrow$ syndrome} (top-right), \emph{syndrome $\rightarrow$ magnitude} (bottom-left), and \emph{syndrome $\rightarrow$ syndrome} (bottom-right). Each of these regions exhibits different behaviors. Notably, the \emph{syndrome $\rightarrow$ syndrome} attention is consistently strong, indicating that the model relies heavily on the syndrome, in addition, the \emph{magnitude $\rightarrow$ syndrome} attention also remains relatively unchanged regardless of the presence of errors, suggesting that the model treats the syndrome as a reference for interpreting the magnitudes, rather than vice versa. Furthermore, when no error is present, the attention in both the \emph{magnitude $\rightarrow$ magnitude} and \emph{syndrome $\rightarrow$ magnitude} regions is low. This implies that the model has learned to infer the presence or absence of errors primarily from the syndrome. However, when an error is present, there is a clear increase in attention across the corresponding column, indicating that the model has learned to examine the entire parity-check line to locate and assess potential errors. In previous analysis  on ECCT \citep{park2024crossmpt} the \emph{magnitude$\rightarrow$magnitude} and \emph{syndrome$\rightarrow$syndrome} relations were less significant leading to the design of the mask in CrossMPT. This analysis shows that ECCM is able to leverage those relation in contrast with previous works.

\section{Limitations and Broader Impacts}
\textbf{Limitations:} While the proposed ECCM decoder demonstrates strong empirical performance and competitive inference efficiency, several limitations should be noted. First, the model architecture, although designed to generalize across code families, was primarily tested on standard benchmarks with moderate block lengths. Its generalization to very long block codes or non-binary codes remains unverified and may require architectural scaling or retraining. Second, while the hybrid Mamba–Transformer structure improves efficiency over attention-only models, the total model complexity remains non-trivial, and resource-constrained environments (e.g., edge devices) may still face deployment challenges. \textbf{Broader Impacts:} Error correction codes are foundational to reliable communication and data storage. The proposed ECCM method improves both the speed and accuracy of decoding. Accuracy improvements can benefit a wide range of technologies, with deep-space transmissions being a notable example, while speed gains may enable learned decoders in real-time systems. However, the black-box nature of learned decoders like ECCM may pose challenges in safety-critical applications where certifiability and interpretability are essential.

\section{Conclusions}

We introduced ECCM, a hybrid Mamba–Transformer decoder for linear error correction codes. By combining Mamba’s efficient sequential modeling with the global context modeling of Transformers, and incorporating parity-check-aware masking and progressive supervision, ECCM achieves state-of-the-art accuracy while maintaining low and improving inference speed. Experimental results across multiple code families demonstrate consistent improvements over existing neural decoders. These findings highlight the potential of hybrid architectures for real-time, high-accuracy decoding, and open the door to further exploration of structured neural models in communication systems.

\bibliographystyle{plain} 
\bibliography{aac}

\appendix

\end{document}